\documentclass[aps,prl,twocolumn,superscriptaddress,amssymb,showpacs]{revtex4}

\usepackage{graphicx}

\begin{document}

%\preprint{}

\title{Periodically-modulated entangled light}



\author{H.~H.~Adamyan}
\email[]{adam@unicad.am}
%\thanks{}
\affiliation{Yerevan State University, A. Manookyan 1, 375049,
Yerevan, Armenia} \affiliation{Institute for Physical Research,
National Academy of Sciences,\\Ashtarak-2, 378410, Armenia}

\author{G.~Yu.~Kryuchkyan}
%\thanks{}
\affiliation{Yerevan State University, A. Manookyan 1, 375049,
Yerevan, Armenia} \affiliation{Institute for Physical Research,
National Academy of Sciences,\\Ashtarak-2, 378410, Armenia}


%\keywords{}

\begin{abstract}
We propose periodically-modulated entangled states of light and
show that they can be generated in two experimentally feasible
schemes of nondegenerate optical parametric oscillator (NOPO): (i)
driven by continuously modulated pump field; (ii) under action of
a periodic sequence of identical laser pulses. We show that the
time-modulation of the pump field amplitude essentially improves
the degree of continuous-variable entanglement in NOPO. We develop
semiclassical and quantum theories of these devices for both
below- and above-threshold regimes. Our analytical results are in
well agrement with numerical simulation and support a concept of
time-modulated entangled states.
\end{abstract}

\maketitle

Continuous-variable (CV) entangled states of light beams provide
excellent tools for testing the foundations of quantum physics and
arouse growing interest due to apparent usefulness as a promising
technology in quantum information and communication protocols
\cite{Braunstein, Furusawa}. The efficiency of quantum information
schemes significantly depends on the degree of entanglement. On
the other hand, in the majority of real applications bright light
beams are required. It is therefore highly desirable to elaborate
reliable sources of light beams having the mentioned properties.
The recent development of CV quantum information is stipulated
mainly by preparation of  EPR (Einstein-Podolsky-Rosen) entangled
states, which particularly can be generated by a nondegenerate
parametric amplifier \cite{Reid, Kimble}. However, up to now the
generation of bright light beams with high degree of CV
entanglement meets serious problems.

The analysis of quantum communication protocols is very easy in
terms of information transfers which can be effectively performed
for communication schemes operating mainly in a pulsed regime
\cite{Grosshans, Wenger, Wenger1}. In this regime it is possible
to manipulate individually each quantum state involved in the
information exchange. This statement has emerged recently and
efficient setups have been proposed for generation and
characterization of quadrature-squeezed pulses \cite{Wenger} as
well as quadrature-entangled pulses \cite{Wenger1} in time-domain
in addition to many other experiments performed in the frequency
domain \cite{Bowen}. In spite of these developments, an important
issue for time-resolved communication protocols is to investigate
CV entanglement for various time-modulated regimes.

As a realization of this program, in this Letter we propose and
investigate the time-modulated entangled states generated in two
schemes of NOPO: (i) driven by continuously modulated pump field;
(ii) under action of a periodic sequence of identical laser
pulses. We stress that these schemes are experimentally feasible
and, that is very remarkable, provide highly effective mechanism
for improvement of the degree of CV entanglement, even in the
presence of dissipation and cavity induced feedback.

CV entangling resources are usually analyzed as a two-mode
squeezing through the variances of the quadrature amplitudes.  In
NOPO, under a continuous, monochromatic pump, the integral
squeezing, which characterizes the entanglement, reaches only
$50\%$ relative to the level of vacuum fluctuations, if the pump
field intensity is close to the generation threshold \cite{Levon}.
As we show below, application of pump laser fields with
periodically-varying amplitudes allows qualitatively improve the
situation, i.e. to go beyond the limit $50\%$, that indicates a
high degree of quadrature entanglement obeying the condition of
EPR-like paradox criterion \cite{Reid}.

We develop quantum theories of these devices for below- and
above-threshold regimes concluding that such achievement takes
place for both operational regimes of NOPO. Noted, that CV
entanglement for ordinary NOPO above threshold have already been
established as theoretically as well experimentally. EPR
entanglement in NOPO above threshold was proposed in \cite{Reid}
and its strong consideration has recently been given in
\cite{Levon}. EPR correlation and squeezing for NOPO above
threshold experimentally confirmed in \cite{Feng}. CV entanglement
of phase-locked light beams for both regimes of NOPO was recently
proposed in \cite{PhaseLocked}.

\begin{figure}
\includegraphics[angle=0,width=0.48\textwidth]{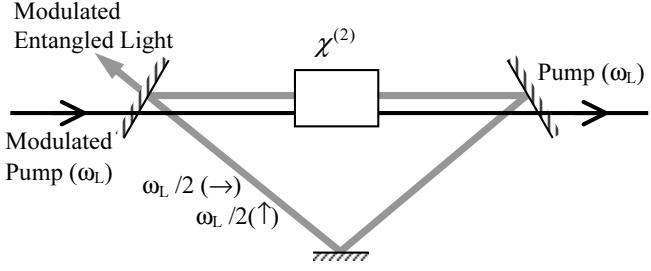}
\caption{The principal scheme of NOPO in a cavity that supports
the pump mode at frequency $\omega_{L}$ and subharmonic modes of
orthogonal polarizations at frequency
$\omega_{L}/2$.}\label{schemeFig}
\end{figure}

We consider a type-II phase-matched NOPO with triply resonant
optical ring cavity under action of pump field with periodically
varying amplitude (see Fig.~\ref{schemeFig}). Below we provide two
concrete examples (i) and (ii) mentioned above. The interaction
Hamiltonian describing both cases within the framework of rotating
wave approximation and in the interaction picture is
\begin{eqnarray}
H &=&i\hbar f\left(t\right)\left( e^{i\left( \Phi _{L}-\omega_{L}
t\right)}a_{3}^{+}-e^{-i\left( \Phi _{L}-\omega_{L} t\right)
}a_{3}\right)\nonumber \\
&&+i\hbar k\left(e^{i\Phi _{k}}a_{3}a_{1}^{+}a_{2}^{+}-e^{-i\Phi
_{k}}a_{3}^{+}a_{1}a_{2}\right),\label{Hamiltonian}
\end{eqnarray}
where $a_{i}$ are the boson operators for cavity modes at the
frequencies $\omega_{i}$. The pump mode $a_{3}$ is driven by an
amplitude-modulated external field at the frequency
$\omega_{L}=\omega_{3}$ with time-periodic, real valued amplitude
$f(t+T)=f(t)$. The constant $ke^{i\Phi _{k}}$ determines an
efficiency of the down-conversion process
$\omega_{L}\rightarrow\frac{\omega_{L}}{2}\left(\uparrow\right)
+\frac{\omega_{L}}{2}\left(\rightarrow\right)$ in $\chi^{(2)}$
medium. We take into account the cavity damping rates $\gamma
_{i}$ of the modes and consider the case of high cavity losses for
the pump mode ($\gamma _{3}\gg \gamma$,
$\gamma_{1}=\gamma_{2}=\gamma$) when the pump mode is eliminated
adiabatically (see Fig.~\ref{schemeFig}). However, in our analysis
we allow for the pump depletion effects. Following the standard
procedure we derive in the positive P-repesentation the stochastic
equations for complex c-number variables $\alpha_{1,2}$ and
$\beta_{1,2}$ corresponding to operators $a_{1,2}$ and
$a^{+}_{1,2}$ for the case of zero detunings:
\begin{eqnarray}
\frac{d\alpha _{1}}{dt}=-(\gamma+ \lambda \alpha_{2}\beta
_{2})\alpha_{1}+\varepsilon\left(t\right)\beta
_{2}+W_{\alpha_{1}}\left(t\right),
\label{alpha1StochEq} \\
\frac{d\beta _{1}}{dt}=-(\gamma+\lambda
\alpha_{2}\beta_{2})\beta_{1}+\varepsilon
\left(t\right)\alpha_{2}+W_{\beta_{1}}\left(t\right).
\label{beta1StochEq}
\end{eqnarray}
Here: $\varepsilon(t)=f(t)k/\gamma_{3}$,
$\lambda=k^{2}/\gamma_{3}$ and equations for $\alpha_{2},
\beta_{2}$ are obtained from (\ref{alpha1StochEq}),
(\ref{beta1StochEq}) by exchanging the subscripts
(1)$\leftrightarrows$(2). Our derivation is based on the Ito
stochastic calculus, and the nonzero stochastic correlations are:
$\langle W_{\alpha_{1}}\left(t\right)W_{\alpha_{2}}
\left(t^{\prime}\right)\rangle = \left(\varepsilon\left(t\right)
-\lambda \alpha _{1}\alpha
_{2}\right)\delta\left(t-t^{\prime}\right)$, $\langle
W_{\beta_{1}}\left(t\right)W_{\beta_{2}}\left(t^{\prime}
\right)\rangle=\left(\varepsilon\left(t\right) -\lambda \beta
_{1}\beta _{2}\right)\delta\left(t-t^{\prime}\right)$. Note, that
while obtaining these equations we used the transformed boson
operators $a_{i}\rightarrow a_{i}exp\left(-i\Phi_{i}\right)$ with
$\Phi_{i}$ being $\Phi_{3}=\Phi_{L}$,
$\Phi_{1}=\Phi_{2}=\frac{1}{2}\left(\Phi_{L}+\Phi_{k}\right)$.
This leads to cancellation of phases at intermediate stages of
calculation.

First, we shall study in general the solution of stochastic
equations in semiclassical treatment, neglecting the noise terms,
for mean photon numbers $n_{j}$ and phases $\varphi_{j}$ of the
modes ($n_{j}=\alpha_{j}\beta_{j}$,
$\varphi_{j}=\frac{1}{2i}ln(\alpha_{j}/\beta_{j})$). An analysis
shows that similar to the standard NOPO, the considered system
also exhibits threshold behavior, which is easily described
through the period-averaged pump field amplitude
$\overline{f(t)}=\frac{1}{T}\int^{T}_{0}f(t)dt$. The
below-threshold regime with a stable trivial zero-amplitude
solution is realized for $\overline{f}<f_{th}$, where
$f_{th}=\gamma\gamma_{3}/k$ is the threshold value. When
$\overline{f}>f_{th}$, the stable nontrivial solution exists with
the following properties. First, as for usual NOPO, the phase
difference is undefined due to the phase diffusion, while the sum
of phases is equal to $\varphi_{1}+\varphi_{2}=2\pi m$. The mean
photon numbers for subharmonic modes
$n_{oi}=\left<a_{i}^{+}a_{i}\right>=\left|\alpha_{i}\right|^{2}$
are equal one to the other ($n_{01}=n_{02}=n_{0}$) due to the
symmetry of the system, $\gamma_{1}=\gamma_{2}=\gamma$. The
straightforward calculations lead to the following result for
over-transient regime
\begin{equation}
n_{0}^{-1}(t)=2\lambda\int^{0}_{-\infty}exp\left(2\int^{\tau}_{0}
\left(\varepsilon\left(t^{\prime}+t\right)-\gamma\right)dt^{\prime}\right)d\tau.\label{asymptoticSolution}
\end{equation}
Note, that $n_{0}(t)$ is a periodic function of time.

To characterize the CV entanglement we address to both the
inseparability criterion \cite{Simon} and the EPR paradox
criterion \cite{Reid}. These criteria could be quantified by
analyzing the variances $V_{-}=V\left(X_{1}-X_{2}\right)$ and
$V_{+}=V\left(Y_{1}+Y_{2}\right)$ in the terms of the quadrature
amplitudes of two modes
$X_{k}=X_{k}\left(\Theta_{k}\right)=\frac{1}{\sqrt{2}}
\left(a^{+}_{k}e^{-i\Theta_{k}}+a_{k}e^{i\Theta_{k}}\right)$, $
 Y_{k}=X_{k}\left(\Theta_{k}-\frac{\pi}{2}\right),\;\left(k=1,2\right)$,
where $V(x)=\left<x^{2}\right>-\left<x\right>^{2}$ is a denotation
of the variance. The inseparability criterion, or weak
entanglement criterion reads as $V_{+}+V_{-}<2$, and due to the
mentioned symmetries is reduced to the following form
$V=V_{+}=V_{-}<1$, while for the product of variances this
criterion has the form $V_{+}V_{-}=V^{2}<1$. The strong CV
entanglement criterion shows that when the inequality
$V_{+}V_{-}<1/4$ is satisfied, there arises an EPR-like paradox.
We consider here the time-dependent output variances, which can be
recorded by time-resolved homodyne detection \cite{Wenger,
Wenger1}. These quantities will be expressed through the
stochastic variables and will be calculated in a linear treatment
of quantum fluctuations. Restoring the previous phase structure of
intracavity interaction, we obtain that $V_{+}=V_{-}=V$ and
\begin{eqnarray}
V=1+\left<\alpha_{1}\beta_{1}\right>+\left<\alpha_{2}\beta_{2}\right>
-\left<\alpha_{1}\alpha_{2}\right>e^{i\Theta}-\left<\beta_{1}\beta_{2}\right>e^{-i\Theta},\label{VarianceWithPhases}
\end{eqnarray}
where $\Theta=\Theta_{1}+\Theta_{2}+\Phi_{L}+\Phi_{k}$.

To this end, it is convenient to use the following moments of
stochastic variables
$\left<n_{+}\right>=\left<\alpha_{1}\beta_{1}\right>
+\left<\alpha_{2}\beta_{2}\right>$,
$\left<R\right>=\left<\left(\alpha_{1}-\beta_{2}\right)\left(\beta_{1}-\alpha_{2}\right)\right>$,
$\left<Z\right>=\left<\left(\alpha_{1}\beta_{1}-\alpha_{2}\beta_{2}\right)^{2}\right>
+\left<\alpha_{1}\beta_{1}\right>+\left<\alpha_{2}\beta_{2}\right>$.
As can be seen, the possible minimal level of variance, realized
under appropriate selection of phases
$\Theta_{1}+\Theta_{2}=-\Phi_{L}-\Phi_{k}$ in formula
(\ref{VarianceWithPhases}), is expressed as
$V(t)=1+\left<R(t)\right>$. Using It\^{o} rules for changing the
stochastic variables, we obtain from (\ref{alpha1StochEq}),
(\ref{beta1StochEq})
\begin{eqnarray}
\frac{d}{dt}\left<n_{+}\right>&=&\left(2\varepsilon\left(t\right)-2\gamma-\lambda\right)\left<n_{+}\right>-\lambda
\left<n_{+}^{2}\right>\nonumber\\
&-&2\varepsilon\left(t\right)\left<R\right>+\lambda
\left<Z\right>,\label{stochasticEquationForNplus}\\
\frac{d}{dt}\left<R\right>&=&-\left(2\varepsilon\left(t\right)
+2\gamma+\lambda\right)\left<R\right>-\lambda
\left<n_{+}R\right>\nonumber\\
&-&2\varepsilon\left(t\right)+\lambda
\left<Z\right>,\label{stochasticEquationForRminus}\\
\frac{d}{dt}\left<Z\right>&=&-4\gamma\left<Z\right>
+2\gamma\left<n_{+}\right>.\label{stochasticEquationForNminus2}
\end{eqnarray}
From Eq.(\ref{stochasticEquationForNminus2}) $\left<Z\right>$ can
be expressed as a function of $\left<n_{+}\right>$. Substituting
this expression into (\ref{stochasticEquationForNplus}),
(\ref{stochasticEquationForRminus}) we get the following equations
which are convenient for the perturbative analysis of quantum
fluctuations
\begin{eqnarray}
\frac{d}{dt}\left<n_{+}\right>&=&\left(2\varepsilon\left(t\right)-2\gamma-\lambda\right)\left<n_{+}\right>-\lambda
\left<n_{+}^{2}\right>-2\varepsilon\left(t\right)\left<R\right>\nonumber\\
&+&2\gamma\lambda\int_{-\infty}^{t}e^{4\gamma\left(\tau-t\right)}{\left<n_{+}\left(\tau\right)\right>d\tau},\label{stochasticEquationForNplusSubstituted}\\
\frac{d}{dt}\left<R\right>&=&-\left(2\varepsilon\left(t\right)
+2\gamma+\lambda\right)\left<R\right>-\lambda
\left<n_{+}R\right>\nonumber\\
&-&2\varepsilon\left(t\right)+2\gamma\lambda\int_{-\infty}^{t}e^{4\gamma\left(\tau-t\right)}{\left<n_{+}\left(\tau\right)\right>d\tau}.\label{stochasticEquationForRminusSubstituted}
\end{eqnarray}
First, we consider the above-threshold regime linearizing quantum
fluctuations around the stable semiclassical solutions:
$\left<n_{+}\right>=n_{10}+n_{20}+\left<\delta
n_{+}\right>=2n_{0}+\left<\delta n_{+}\right>$,
$\left<R\right>=R^{0}+\left<\delta R\right>=\left<\delta
R\right>$, $\left<n_{+}R\right>=2n_{0}\left<\delta R\right>$,
$\left<n_{+}^{2}\right>=4n_{0}\left<\delta n_{+}\right>$, where it
is assumed that $n_{10}=n_{20}=n_{0}(t)$,
$\varphi_{1}+\varphi_{2}=2\pi k$, and hence $R^{0}=0$. Note, that
in the current experiments the ratio of nonlinearity to dumping is
small, $k/\gamma\ll 1$ (typically $10^{-4}$ or less), and hence
$\lambda/\gamma=k^{2}/\left(\gamma\gamma_{3}\right)\ll 1$ is the
small parameter of the theory. Therefore, the zero order terms in
the above expansion correspond to a large classical field of the
order $\gamma/\lambda$ in accordance with
Eq.(\ref{asymptoticSolution}), while the next terms describing the
quantum fluctuations are of the order of $1$. On the whole,
combining the procedure of linearization with $\lambda/\gamma\ll
1$ approximation we get a linear equation for the variance
$V(t)=1+\left<\delta R\right>$
\begin{eqnarray}
\frac{d}{dt}V\left(t\right)&=&-2\left(\gamma+\varepsilon\left(t\right)+\lambda
n_{0}\left(t\right)\right)V\left(t\right) + 2\lambda n_{0}\left(t\right)\nonumber\\
&+& 2\gamma
+4\gamma\lambda\int_{-\infty}^{t}e^{4\gamma\left(\tau-t\right)}{n_{0}(\tau)d\tau},\label{linearizedV}
\end{eqnarray}
with the following periodic asymptotic solution
\begin{eqnarray}
V\left(t\right)=2\int_{-\infty}^{t}{\exp\left(-2\int_{\tau}^{t}{\left(\gamma
+\varepsilon\left(t^{\prime}\right)+\lambda n_{0}\left(t^{\prime
}\right)\right)dt^{\prime}}\right)}\times\nonumber\\
\left[\gamma+\lambda
n_{0}\left(\tau\right)+2\gamma\lambda\int_{-\infty}^{\tau}e^{4\gamma\left(\tau^{\prime}-\tau\right)}
n_{0}(\tau^{\prime})d\tau^{\prime}\right]d\tau.
\label{asymtoticSolutionV}
\end{eqnarray}
The analysis of the below-threshold regime is more simple and
leads to formula (\ref{asymtoticSolutionV}) with $n_{0}=0$.

Let us now consider the output behavior of NOPO assuming that all
losses occur through the output coupler (see, Fig. 1). In this
case the output fields are
$\Phi^{out}_{i}(t)=\sqrt{2\gamma}a_{i}(t),~(i=1,2)$ and
$\Phi^{out}_{3}(t)=\sqrt{2\gamma_{3}}a_{3}(t)-\Phi^{in}_{3}(t)$,
while the output measured time-dependent variances are
$V_{+}^{out}=V_{-}^{out}=V^{out}(t)=2\gamma V(t)$ and the mean
photon number is $n^{out}(t)=2\gamma n_{0}(t)$. We present below
applications of these results to two concrete schemes.

(i) \emph{Model of continuously-modulated NOPO}. The corresponding
scheme (Fig.~\ref{schemeFig}) involves pump field with the
modulated amplitude $f(t)=f_{0}+f_{1}cos(\delta t)$, where
$\delta$ is the modulation frequency, $\delta\ll\omega_{L}$. Such
modulation may be realized by the standard methods, particularly,
for NOPO driven by a polychromatic pump field with central
frequency $\omega_{L}$ and two satellites $\omega_{L}+\delta$,
$\omega_{L}-\delta$. In the last case the Hamiltonian of this
system is indeed given by (\ref{Hamiltonian}) and $f_{0}$ and
$f_{1}$ are the amplitudes of the central component and the
satellites of the pump field. In above threshold,
$\overline{f}=f_{0}>f_{th}$, the photon number
(\ref{asymptoticSolution}) reads as
\begin{eqnarray}
n_{0}^{-1}(t)&=&2\lambda\int^{0}_{-\infty}\exp\left(2\gamma\tau\left(\frac{\overline{f}}{f_{th}}-1\right)\right)\times\nonumber\\
&&\exp\left(\frac{2\gamma f_{1}}{\delta
f_{th}}\left[\sin\left(\delta\left(
t+\tau\right)\right)-\sin\left(\delta t\right)\right]
\right)d\tau.\label{asymptoticSolutionForNForHarmonicModulation}
\end{eqnarray}
\begin{figure}
\includegraphics[angle=-90,width=0.235\textwidth]{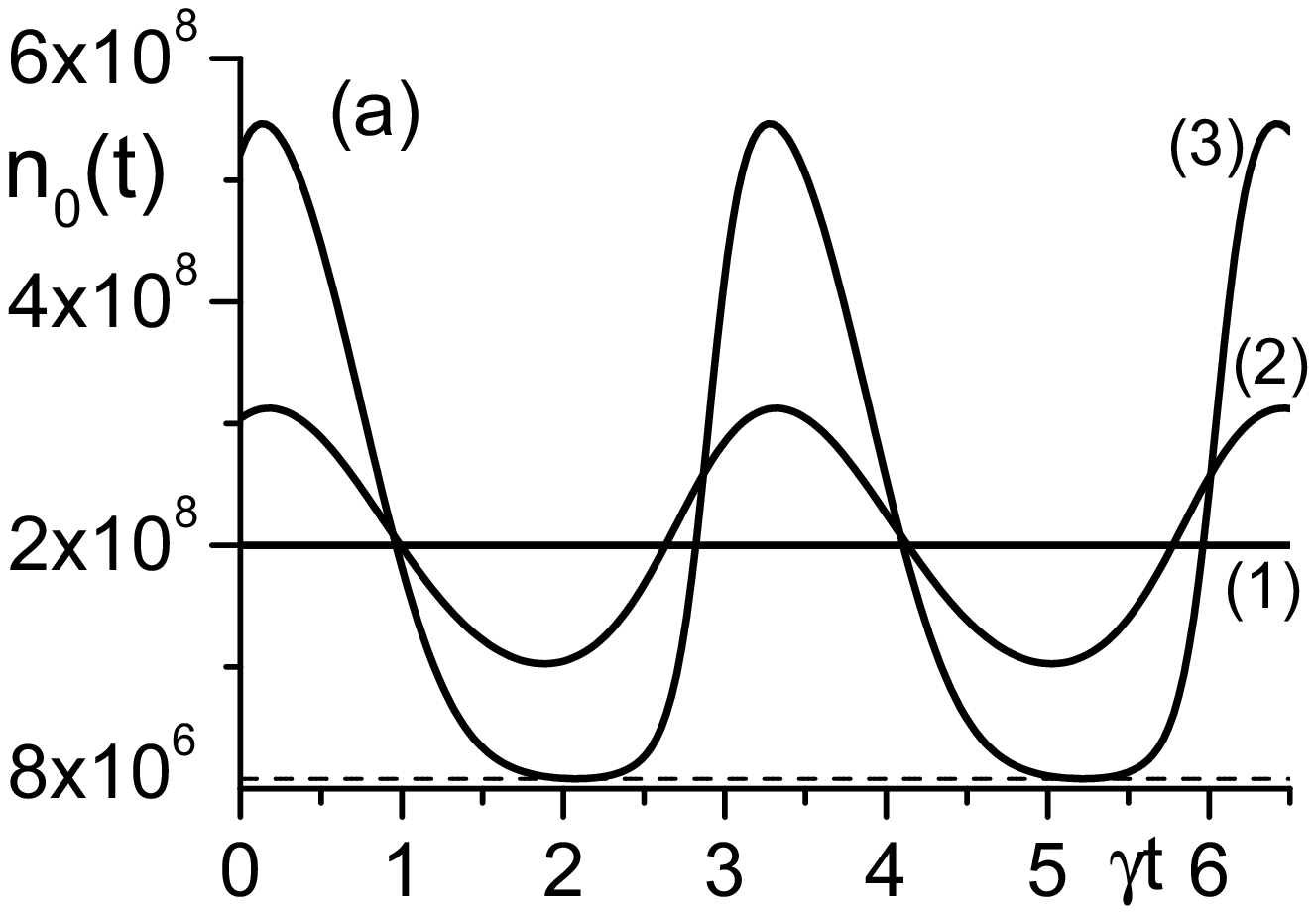}
\includegraphics[angle=-90,width=0.235\textwidth]{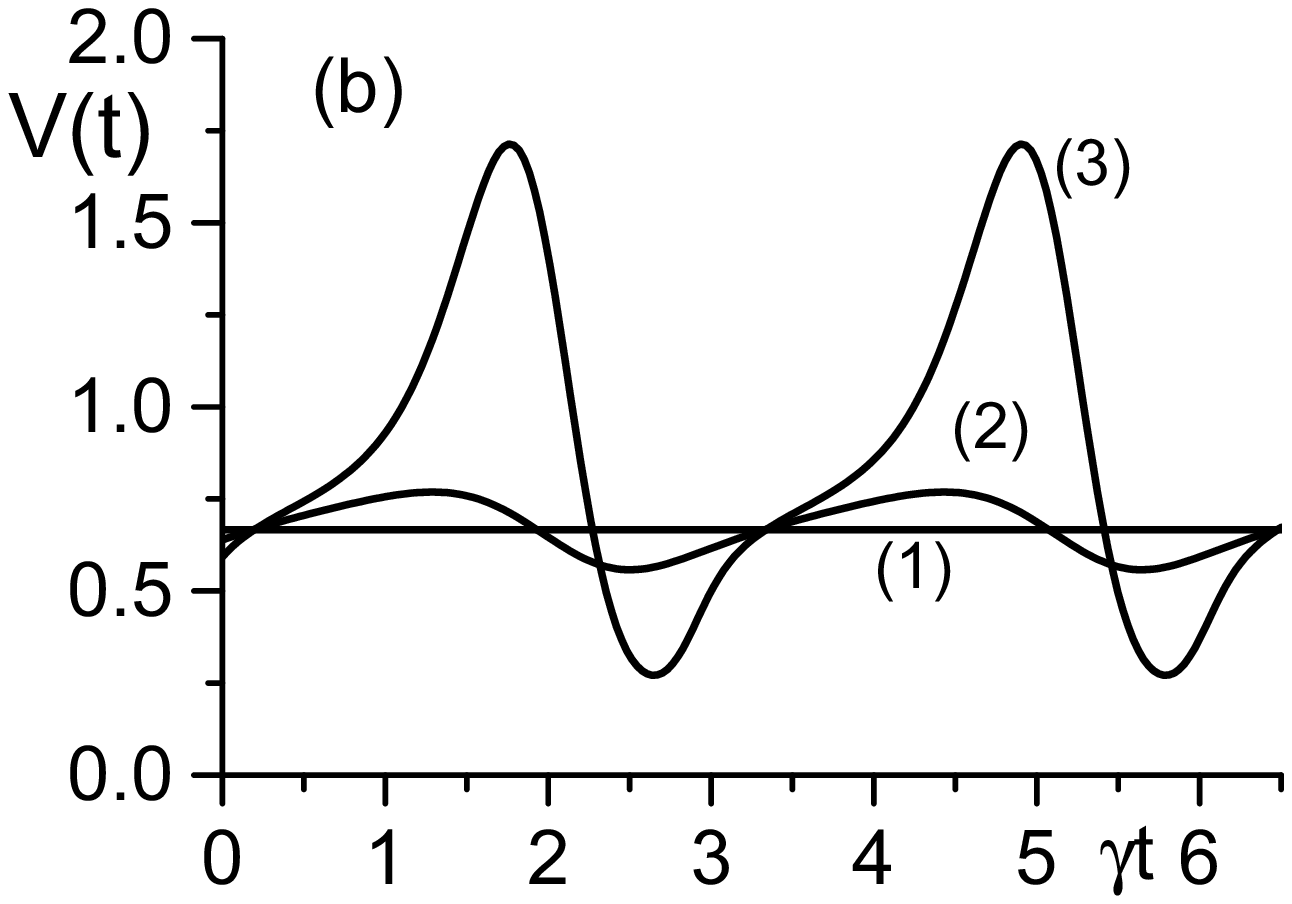}
\caption{Mean photon number (a) and the variance
$V(t)=V^{out}(t)/2\gamma $ (b) versus dimensionless time for the
parameters: $k/\gamma=5\cdot10^{-4}$, $\gamma_{3}/\gamma=25$,
$\delta/\gamma=2$, $\overline{f}=3f_{th}$: $f_{1}=0$ (curve 1),
$f_{1}=0.4\overline{f}$ (curve 2) and $f_{1}=1.2\overline{f}$
(curve 3).} \label{Photon_fig}
\end{figure}
\begin{figure}
\includegraphics[angle=0,width=0.235\textwidth]{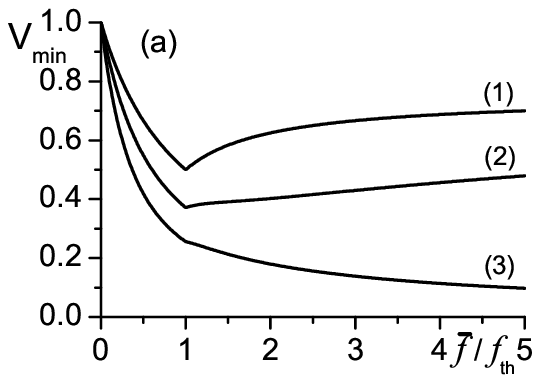}
\includegraphics[angle=0,width=0.235\textwidth]{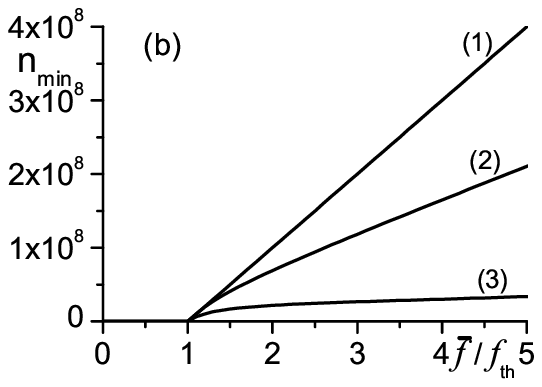}
\caption{The minimum level of the variance (a) and the mean photon
number at the points of minima of the variance (b) versus
$\overline{f}/f_{th}$ for three levels of modulation: $f_{1}=0$
(curve 1), $f_{1}=0.75\overline{f}$ (curve 2) and
$f_{1}=2\overline{f}$ (curve 3). The parameters are:
$k/\gamma=5\cdot10^{-4}$, $\gamma_{3}/\gamma=25$,
$\delta/\gamma=2$.} \label{Vmin_fig}
\end{figure}
This result is illustrated in Fig.~\ref{Photon_fig}a for the
different levels of modulation and for $f_{1}=0$ reaches to the
standard result $n^{out}=2\gamma(f_{0}-f_{th})/k$. Let us turn to
study the entanglement on the formula (\ref{asymtoticSolutionV}),
which for $f_{1}=0$ also coincides with an analogous one for the
ordinary NOPO. Typical results for
$\varepsilon\left(t\right)=\frac{k}{\gamma_{3}}\left(f_{0}+f_{1}\cos\left(\delta
t\right)\right)$ are presented in Fig.~\ref{Photon_fig}b for the
above-threshold regime. The variance is seen to show a
time-dependent modulation with a period $2\pi/\delta$. The drastic
difference between the degree of two-mode squeezing/entanglement
for modulated and stationary dynamics is also clearly seen in
Fig.~\ref{Photon_fig}b. The stationary variance (curve 1) near the
threshold having a limiting squeezing of $0.5$ (see also
Fig.~\ref{Vmin_fig}a, curve 1) is bounded by quantum
inseparability criterion $V<1$, while the variance for the case of
modulated dynamics obeys the EPR criterion $V^{2}<1/4$ of strong
CV entanglement for definite time intervals. The minimum values of
the variance $V_{min}=V\left(t_{m}\right)=V^{out}(t_{m})/2\gamma$
and corresponding photon numbers $n_{min}=n_{0}(t_{m})$ of
Figs.~\ref{Photon_fig} at fixed time intervals $t_{m}=t_{0}+2\pi
m/\delta$, ($m=0,1,2...$) are shown in Figs.~\ref{Vmin_fig}. As it
is expected, the degree of EPR entanglement increases with ratio
$f_{1}/\overline{f}$. The production of strong entanglement occurs
for the period of modulation comparable with the characteristic
time of dissipation, $\delta\approx\gamma$ and dissapears for
asymptotic cases of slow ($\delta\ll\gamma$) and fast
($\delta\gg\gamma$) modulations.

\begin{figure}
\includegraphics[angle=-90,width=0.235\textwidth]{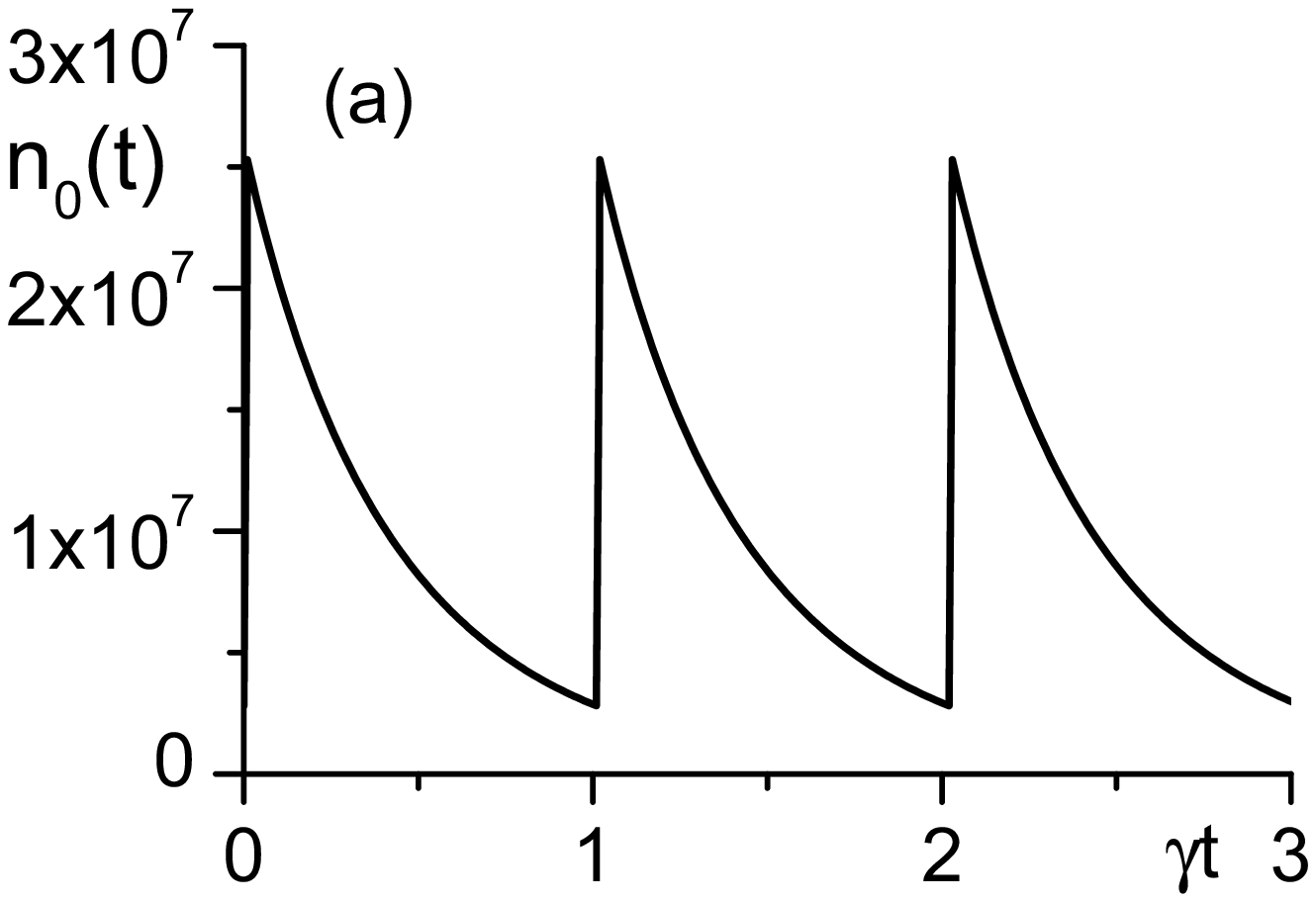}
\includegraphics[angle=-90,width=0.235\textwidth]{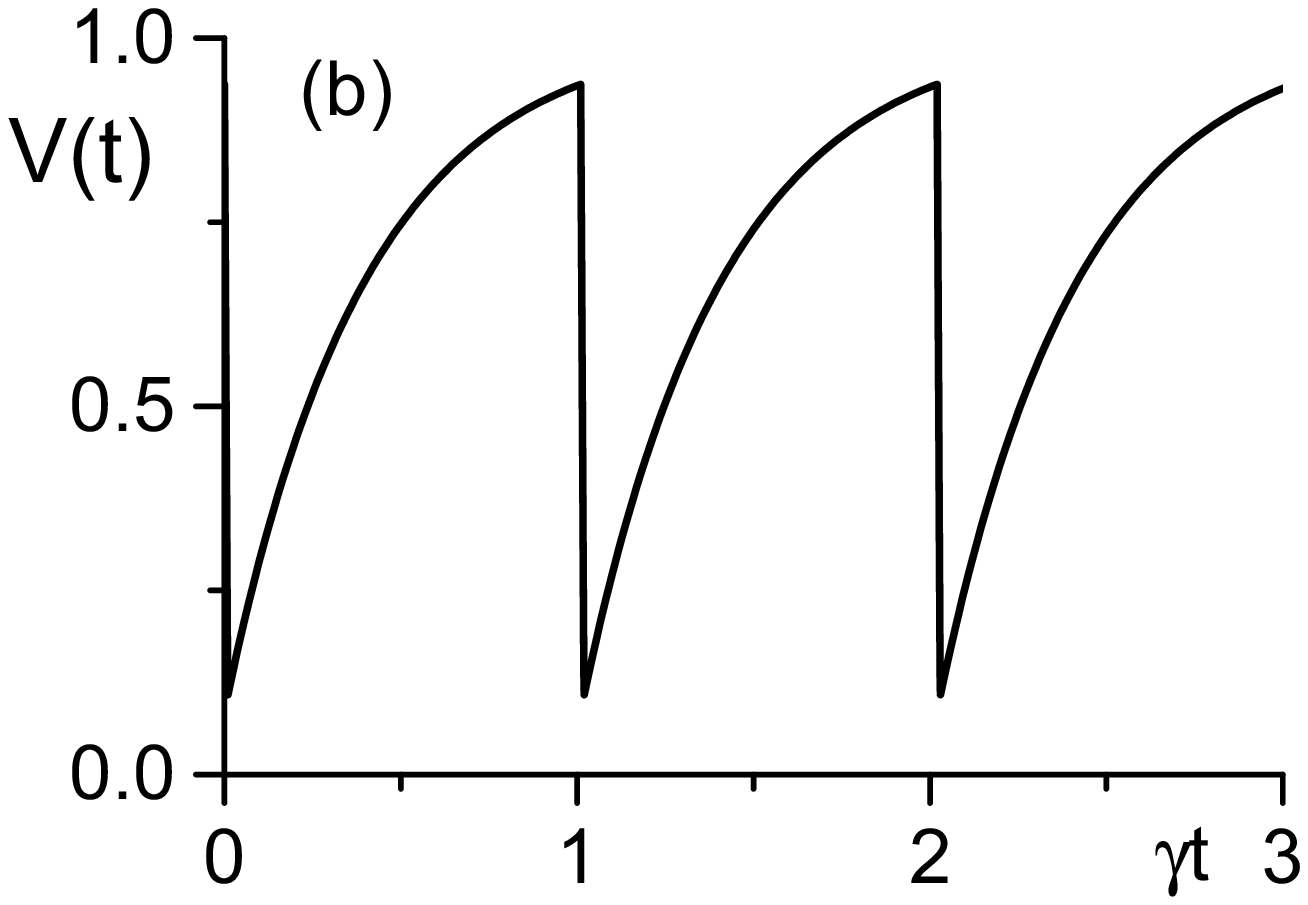}
\caption{Mean photons number (a) and the variance (b) versus
dimensionless time for the parameters: $k/\gamma=5\cdot10^{-4}$,
$\gamma_{3}/\gamma=25$, $T_{1}=0.01\gamma^{-1}$,
$T_{2}=\gamma^{-1}$, $\overline{f}=1.1f_{th}$.}
\label{SigmaPhoton_fig}
\end{figure}

(ii) \emph{Model of periodically pumped NOPO}. We turn now to the
scheme of Fig.~\ref{schemeFig} subjected by a periodic sequence of
identical laser pulses. We consider a rectangular form of the
pulses of the duration $T_{1}$ assuming that $T_{1}$ is much less
than the interval $T_{2}$ between the pulses. Period averaged pump
field amplitude $\overline{f}=f_{L}T_{1}/(T_{1}+T_{2})$, where
$f_{L}$ is the highness of laser pulses, and hence the
above-threshold regime is realized if
$f_{L}T_{1}>\frac{\gamma\gamma_{3}}{k}(T_{1}+T_{2})$. The mean
photon numbers and the variance $V(t)$ are calculated on the
formulas (\ref{asymptoticSolution}) and
(\ref{asymtoticSolutionV}). The predictions of the numerical
calculations are shown on Figs.~\ref{SigmaPhoton_fig} for one of
the preferable regimes (for typical $\gamma=10^{6}s^{-1}$,
$T_{1}=10^{-8}s$ and the repetition rate $T_{2}^{-1}=1MHz$). It is
clearly evident from Fig.~\ref{SigmaPhoton_fig}a the mean photon
number increases during laser pulses and decays during the
interval $T_{2}$ between pulses due to dissipation in the cavity.
One can conclude from Fig.~\ref{SigmaPhoton_fig}b that the weak
entanglement criterion $V<1$, is fulfilled for any time intervals.
However, we have also found remarkable result that the variance
goes below the inseparability level of $0.5$ in the ranges of
maximal photon numbers, for appropriate chosen parameters. It has
occurred for non-stationary regime, if $T_{1}$ is enough shorter
than the relaxation time and hence the dissipative effects in
modes dynamics are still unessential. We illustrate these results
by calculation of the minimum values $V_{min}$. Considering for
simplicity NOPO below and near the threshold and assuming
$T_{1}\ll T_{2}$, we get from Eq.~(\ref{asymtoticSolutionV})
\begin{equation}
V_{min}=e^{-2\varepsilon_{L}T_{1}}\frac{1-e^{-2\gamma
T_{2}}}{1-e^{-2\gamma
T_{2}-2\varepsilon_{L}T_{1}}},\label{VminPulsed}
\end{equation}
where $\varepsilon_{L}=f_{L}k/\gamma_{3}$. This formula is in
accordance with the data of Fig.~4b. As we see the degree of EPR
entanglement increases with $\varepsilon_{L}T_{1}$.

It is well known that the linearized theory is applicable only
outside the critical region, although the variance
(\ref{asymtoticSolutionV}) is surprisingly well defined also at
the threshold. As our analysis shows, the condition of the
validity of linear results for the near-threshold regimes reads as
$\left|\overline{f}/f_{th}-1\right|\gg\left(\lambda/\gamma\right)
\exp\left[2(f_{1}/f_{th})(\gamma/\delta)\right]$ for the system
(i), while for (ii) the condition takes more simple form
$\left|\overline{f}/f_{th}-1\right|\gg\left(\lambda/\gamma\right)$.
For typical $\lambda/\gamma\ll 1$, both conditions are fairly easy
to satisfy even for narrow critical ranges. Note, that the
accuracy of our analytical calculations has been verified by the
numerical simulations based on the quantum state diffusion method.

In conclusion, we note that both schemes (i) and (ii) operate
under non-stationary conditions that has a significant impact on
formation of high-degree CV entanglement even in the presence of
dissipation and cavity induced feedback. We stress that the
properties of periodically pulsed entanglement can be widely
controlled via the modulation parameters. We would like to point
out also that time-dependent output variance could be observed by
means of time-resolved homodyne measurements \cite{Wenger,
Wenger1}. We believe that the results obtained are applicable to a
general class of quantum dissipative systems and can serve as a
guide for further studies of entanglement physics in application
to time-resolved quantum information protocols.

\begin{acknowledgments}
Acknowledgments: This work was supported by NFSAT PH 098-02/CRDF
-12052 and ANSEF PS 89-66 grants.
\end{acknowledgments}

\end{document}